\documentclass[aps,preprint]{revtex4}
\usepackage{float}\usepackage{float}\usepackage{float}
\usepackage{epsfig}
\usepackage{float}
\usepackage{amsmath}
\usepackage{epsfig}

\begin{document}
\title
{\bf Low reflection at zero or low-energies in the well-barrier scattering potentials}
\author{Zafar Ahmed$^1$, ~Sachin Kumar$^2$ and Dhruv Sharma$^3$}
\affiliation{$~^1$Nuclear Physics Division,$~^2$Theoretical Physics Section, Bhabha Atomic Research Centre, Mumbai 400085, India\\
$^{3}$Department of Physics, The Max Planck Institute for Gravitational Physics (Albert Einstein Institute), Hannover }
\email{1:zahmed@barc.gov.in, 2:sachinv@barc.gov.in, 3:sharmadhruv@gmail.com}
\date{\today}
\begin{abstract}
\noindent
Probability of reflection $R(E)$ off a finite  attractive scattering potential at zero or low energies  is ordinarily supposed to be 1. However, a fully attractive  potential presents a paradoxical result that $R(0)=0$ or $R(0)<1$, when an effective parameter $q$ of the potential admits special discrete values. Here, we report another class of finite potentials which  are well-barrier (attractive-repulsive) type  and which can be made to possess much less reflection at zero and low energies for a band of low values of $q$. These  well-barrier potentials have only two real turning points for 
$E \in(V_{min}, V_{max})$, excepting $E=0$. We present two exactly solvable and two numerically solved models to confirm this phenomenon. 
\end{abstract}
\maketitle
\section{Introduction}
Ordinarily, the reflection probability $R(E)$ of a particle of  zero (extremely low) energy incident on a one-dimensional potential which converges to zero asymptotically is found to be 1: $R(0)=1$, the single Dirac delta and the square  well potentials are the simplest textbook examples [1]. This observation is also intuitive, for a zero-energy particle the tunnel effect is negligible such that the transmission probability is close to zero. A paradoxical phenomenon that $R(0)=0$ or $R(0)<1$ has been revealed and proved as a threshold anomaly [1-5] for a potential which is at the threshold of binding a state at $E=0.$ This paradoxical result may be understood in terms of wave packet scattering from an attractive potential. A wave packet with zero average kinetic energy, localized to one side of the potential, will spread in both directions. When the low energy components scatter against the potential, they are transmitted and this would appear simply as wave packet spreading preferentially to the  other direction.

First Senn [2] used an attractive double Dirac delta potential (DDDP) to demonstrate this paradoxiacal reflection.  Nogami and Ross [3] separated two cases: symmetric and asymmetric scattering potentials, to show $R(0)=0$ and $R(0) <1)$, respectively. Next Kiers and van Dijk [4] concluded the same in $n-$channel one dimension scattering from symmetric and asymmetric potentials. In these works DDDP has been used as a convenient and amenable model to demonstrate the paradoxical reflection at zero energy.

Recently, the paradoxical reflection has been shown to exist in the most simple square and exponential wells [5] which are symmetric. Low reflection at low energies has also been discussed. The connection of $R(0)=0$ with half bound state (HBS) of these potentials has been discussed well. An HBS can be defined as $n-$node state at $E=0$ such that the wave function satisfies the Neumann boundary condition that $\psi(\pm \infty)=$constant and there exist $n$ number of bound states at energy $E<0.$ In distinction to  bound states, we denote the solitary HBS of a potential as $\psi_*(x)$, see figs. 1(a) and 3. Since these two potential wells are symmetric the result $R(0)<1$ could not be encountered (we got $R(0)=0$). In a crucial comment to Ref. [5], van Dijk and Nogami [6] emphasized the occurrence of $R(0)<1$
in non-symmetric DDDP model. Interestingly, we found this comment [6] useful to visualize asymmetric DDDP ($\lambda \tilde \lambda<0$ in [6]) as a well barrier potential.

In this note we present the phenomenon of low reflection at zero or low energies when the well barrier system possesses $n$-node HBS, interestingly $n$ could also be zero then the HBS will be node less. We present, the exactly solvable DDDP and Scarf II [7] potentials. We also present two numerically solved  models of well and barrier potentials to demonstrate low reflection at low energy and $R(0)<1.$

\section{Two exactly solvable well-barrier systems}
In order to bring out the conditional existence of $R(0)<1$ and its connection with HBS, we first discuss two exactly solvable models: Double Dirac delta potential (DDDP) and Scarf II [7] potentials. 
\subsection{Double Dirac delta well barrier potential}
DDDP [1-5] is depicted in Fig. 1(a) by solid line and it is written as
\begin{equation}
V(x)=-U_1 \delta(x)+U_2 \delta(x-a), U_1,U_2>0
\end{equation}
Let us define $2\mu U_j/\hbar^2=u_j$. This potential is known to have one bound state if $u_1a >1$. The reflection amplitude $r(E)$ of a general DDDP has been derived [3] which for (1) can be written as
\begin{equation}
r(E)=-\frac{2ik(u_1 e^{-ika}-u_2 e^{ika})+2i u_1u_2\sin ka}{(2ik+u_1)(2ik-u_2)e^{-ika}+u_1 u_2e^{ika}}, \quad R(E)=| r(E)|^2.
\end{equation}
$r(0)$ becomes indeterminate $0/0$, but the limit of $|r(E)|^2$ as $E\rightarrow 0$
equals 1. 
\begin{figure}[t]
	\centering
	\includegraphics[width=7 cm,height=5 cm]{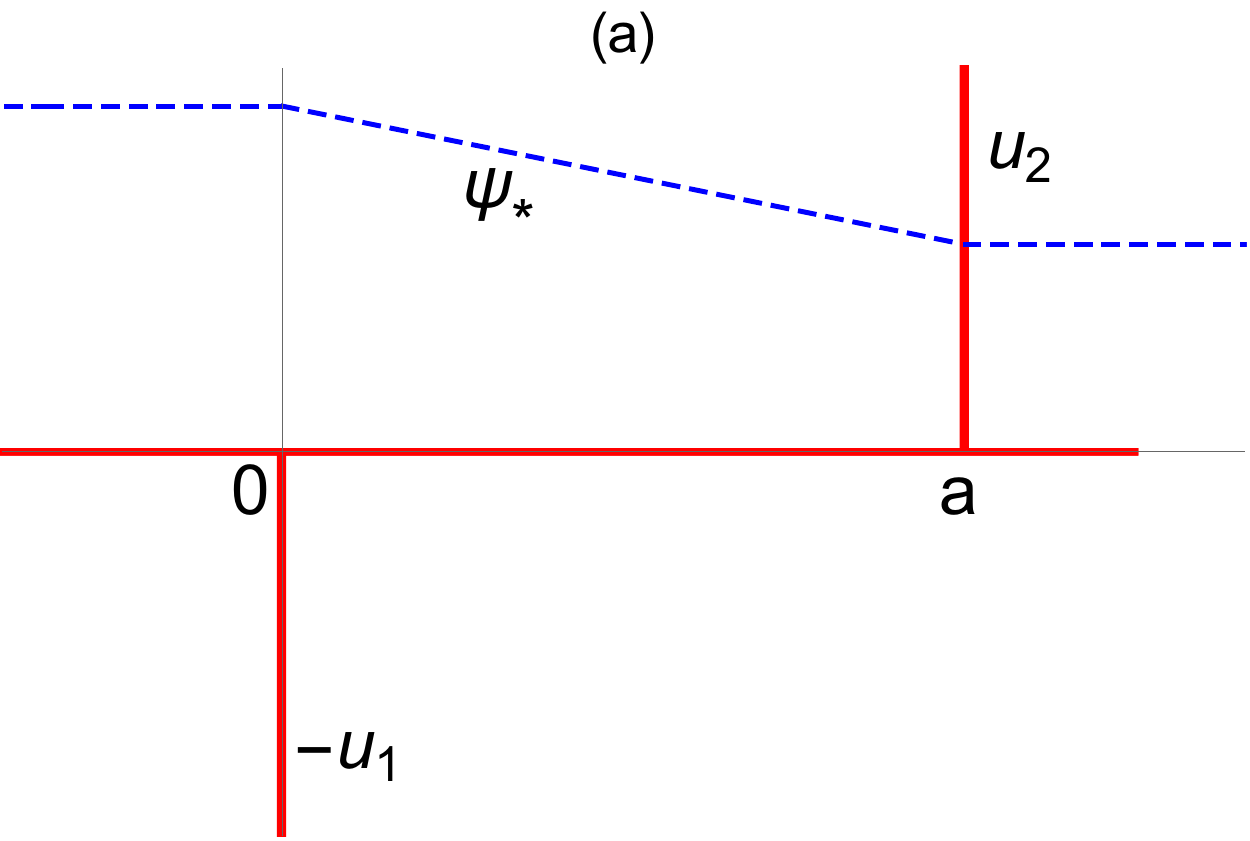}
	\hskip .5 cm
	\includegraphics[width=7 cm,height=5 cm]{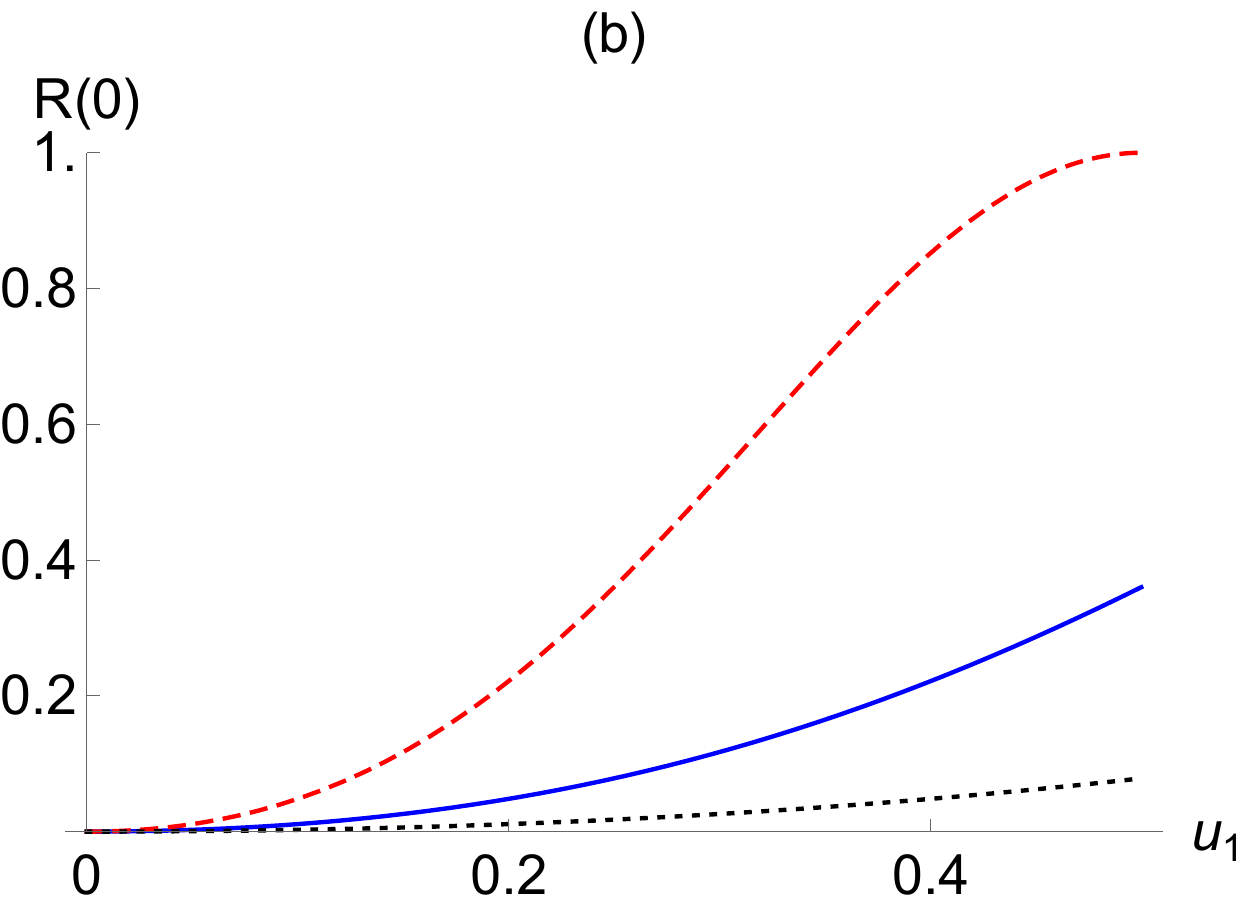}
	%\hskip .5 cm
	%\includegraphics[width=3 cm,height=5 cm]{fig-pdx-1d.pdf}
	\caption{The simplest Dirac delta well-barrier system (solid line) in (a)  and the nodeless half bound state $\psi_*(x)$ at $E=0$ in dashed line, when $1/u_2-1/u_1=a$ (3) is met critically and we get $R(0)<<1$. In (b), $R(0)$ is plotted as a function of $u_1$, when $a$ is kept as 0.5 (dotted), 1 (solid) and 2 (dashed) and $u_2$ satisfies the relation (3). Notice that when the well and barrier are nearer ($a$ is smaller) $R(0)$ is lesser.}
\end{figure} 
The zero energy solution of Schr{\"o}dinger equation can be written piece-wise as $\psi_*(x\le 0){=} A, ~\psi_*({0<x<a}) {=} Bx+C, ~\psi_*(x\ge a){=} D$. Matching these solutions and mis-matching their derivative at $x{=}0$ and $a$, we get $A{=}C, ~B{=}-u_1 A,~ Ba+C{=}D, ~-B{=}u_2 D$. The eliminnant of these linear equations gives the condition for the existence of HBS at $E=0$ in DDDP (1) as
\begin{equation}
u_2=\frac{u_1}{1-u_1a},\quad u_1 a<1, u_2>0.
\end{equation}
Then, under the condition  of half bound state (3), from Eq.(2) we get
\begin{equation}
\lim_{E\rightarrow 0} R(E)=\left | \frac{u_1^2a^2-2u_1 a}{u_1^2a^2-2u_1 a+2}\right |^2 <<1.
\end{equation}
In Fig. 1(b), we show $R(0)$ as a function of $u_1$ for three values of $a=$ 0.5 (dotted), 1 (solid) and 2 (dashed), here $u_2$ satisfies the HBS condition (3). In Fig. 1(a), see the zero energy, 
piece-wise zero-curvature nodeless HBS.
\subsection{The versatile Scarf II well-barrier potential}
The versatile Scarf II potential [7] can be written as
\begin{equation}
V(x)=(s^2-q^2-q)\mbox{sech}^2x+s(2q+1) \mbox{sech}x \tanh x
\end{equation}

The  elegant forms of the reflection and transmission amplitudes of this potential have been
derived in terms of Gamma functions $\Gamma(z)$ of complex arguments. Using various properties these functions we can derive the transmission $T(E)$ and reflection $R(E)$ probabilities as
\begin{eqnarray}
T(E)=\frac{\sinh^2 \pi k  \cosh^2 \pi k}{(\sinh^2 \pi k + \sin^2 \pi q)(\sinh^2 \pi k+ \cosh^2 \pi s)} \nonumber \\
\hspace*{-1cm} R(E)=\frac{\sin^2 \pi q (\sinh^2 \pi k+\cosh^2 \pi s)+\sinh^2 \pi k \sinh^2 \pi s }{(\sinh^2 \pi k + \sin^2 \pi q)(\sinh^2 \pi k+ \cosh^2 \pi s)}
\end{eqnarray}
It is interesting to check that when $q$ is not an integer we get ordinary result that $T(0)=0$ and $R(0)=1$. But if $q$ is an integer, we get
\begin{equation}
\lim_{k\rightarrow 0}T(E)=\lim_{k\rightarrow 0} \frac{\sinh^2 \pi k \cosh^2 \pi k}{\sinh^2 \pi k (\sinh^2 \pi k+\cosh^2 \pi s)}= \mbox{sech}^2 \pi s
\end{equation}
and
\begin{equation}
\lim_{k\rightarrow 0} R(E)=\lim_{k\rightarrow 0} \frac{\sinh^2 \pi k \sinh^2 \pi s}{\sinh^2 \pi k (\sinh^2 \pi k+\cosh^2 \pi s)}=\tanh^2 \pi s <<1.
\end{equation}
\vspace*{.2 cm}
They satisfy $R(0)+T(0)=1$, but $R(0)<T(0)$ (7,8), for small values of $s$, is paradoxical,.
\begin{figure}[t]
	\centering
	\includegraphics[width=7 cm,height=5 cm]{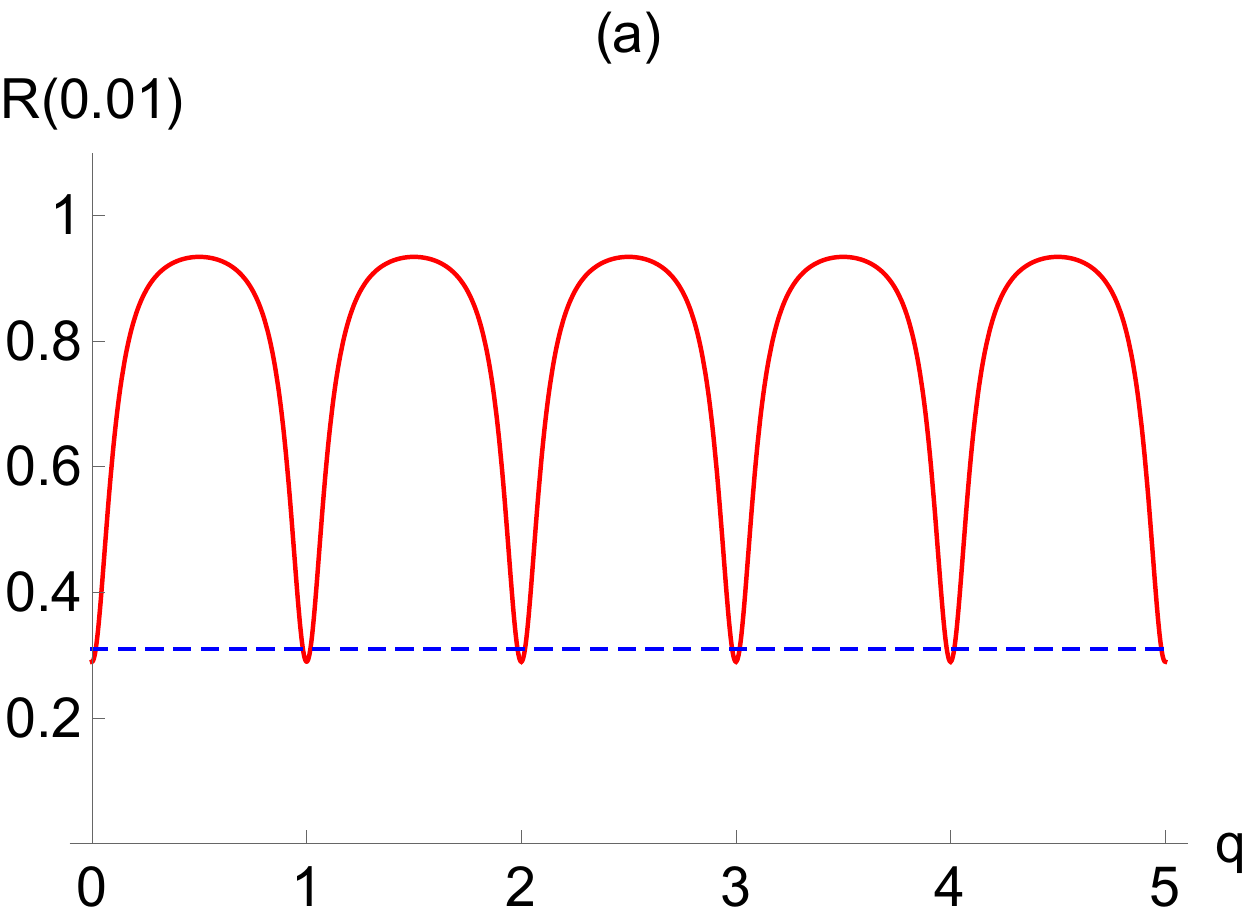}
	\hskip .5 cm
	\includegraphics[width=7 cm,height=5 cm]{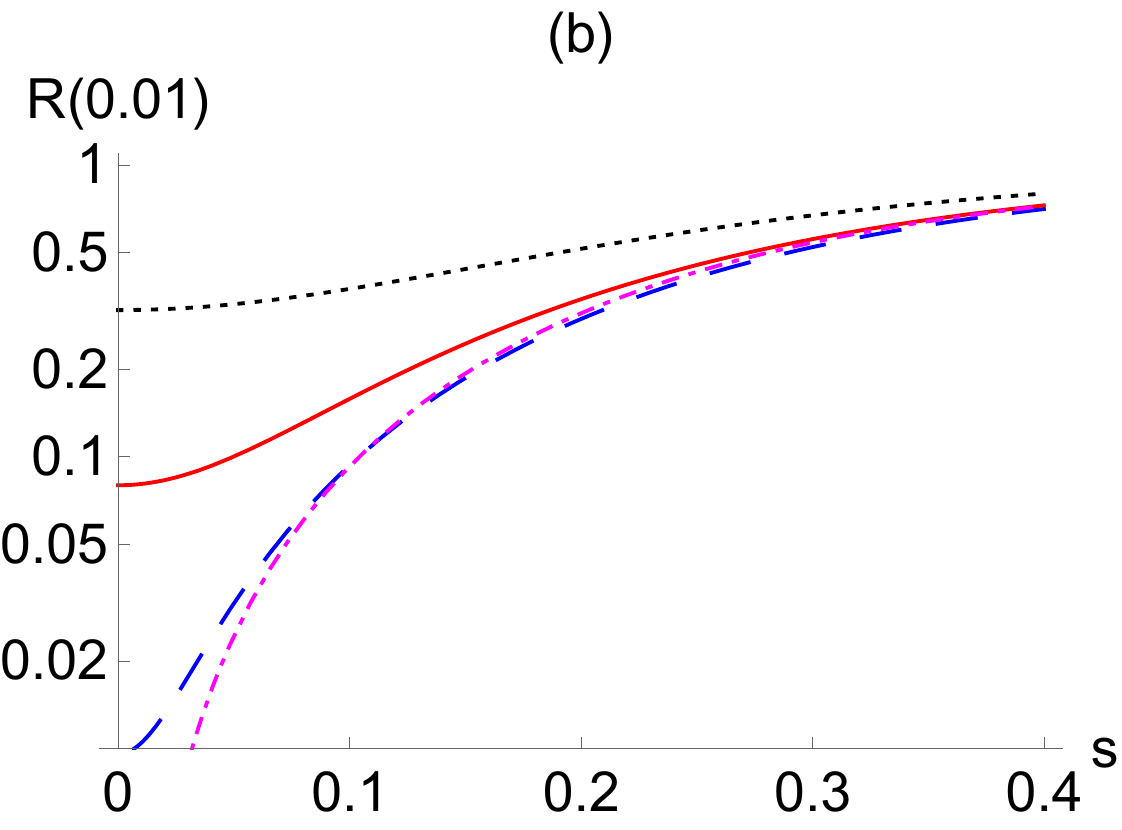}
	%\hskip .5 cm
	\caption{$R(0.01)$ as a function of $q$ for Scarf II potential ($s=0.2$), notice that when $q$ takes values around an integer,  $R(0.01)$ is much less and even lesser than $R(0)=\tanh^2 \pi s$ (dashed line) in (a). In (b) we plot $R(0.01)$ as a function of $s$ for three different values of $q$ close to integers as $1.01$ (solid), $\pm 0.03$ (dashed) and $1.97$ (dotted). Dot-dashed line is $R(0)=\tanh^2 \pi  s$ (8) (when $q$ is an integer). Notice that in a band of small $q$ values $(q<0.1)$, we have $R(0.01)<<1$}
\end{figure}

For this potential the bound state eigenvalues are given as [8]
\begin{equation}
E_n=-(n-q)^2,~ n=0,1,2,3,...[q], ~q \notin I.
\end{equation}
\begin{multline}
\psi_n(x)=i^n (1+y^2(x))^{[(-q/2) \tan^{-1}y(x)]}~ {\cal P}^{(is-q-1/2,-is-1/2)}_n (iy(x)), ~n=0,1,2,..[q]. 
\end{multline}
We find that, if $q$ is an integer then $\psi_{n=q}(x)$ becomes the critical solitary HBS $\psi_*(x)$ of the Scarf II potential (5). Here $y(x)=\sinh x$ and ${\cal P}^{\alpha, \beta}_n(z)$ are Jacobi polynomials.
\begin{figure}[H]
	\centering
	\includegraphics[width=5 cm,height=5 cm]{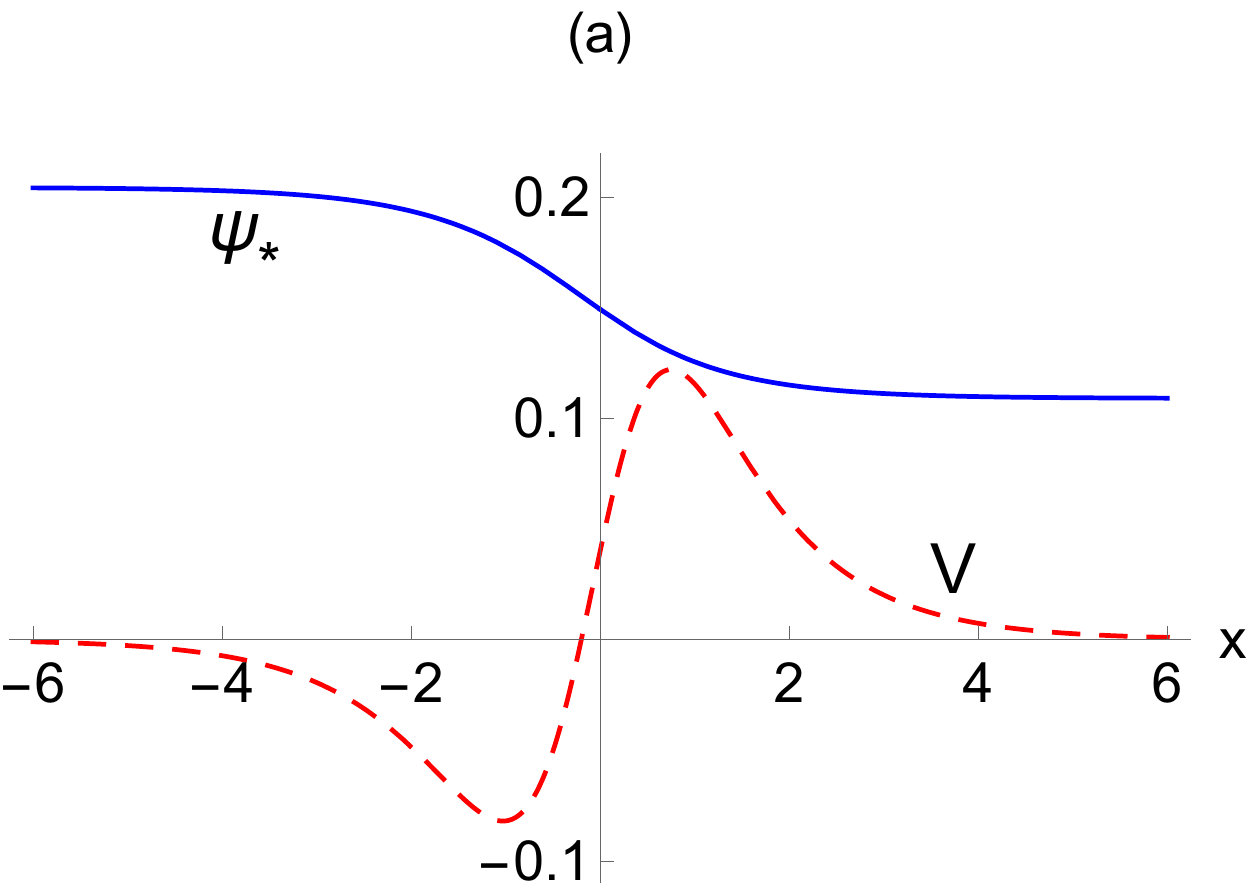}
	\hskip .5 cm
	\includegraphics[width=5 cm,height=5 cm]{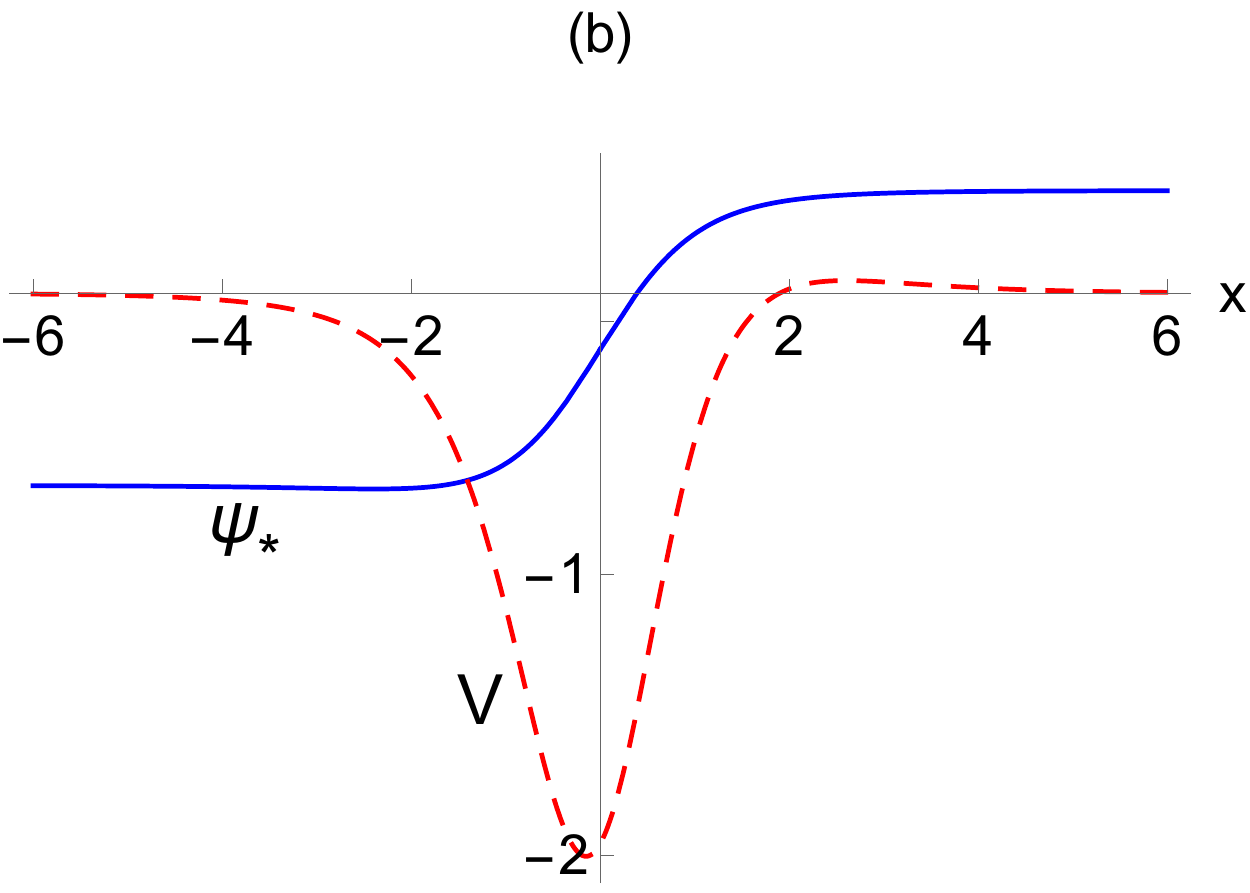}
	\hskip .5 cm
	\includegraphics[width=5 cm,height=5 cm]{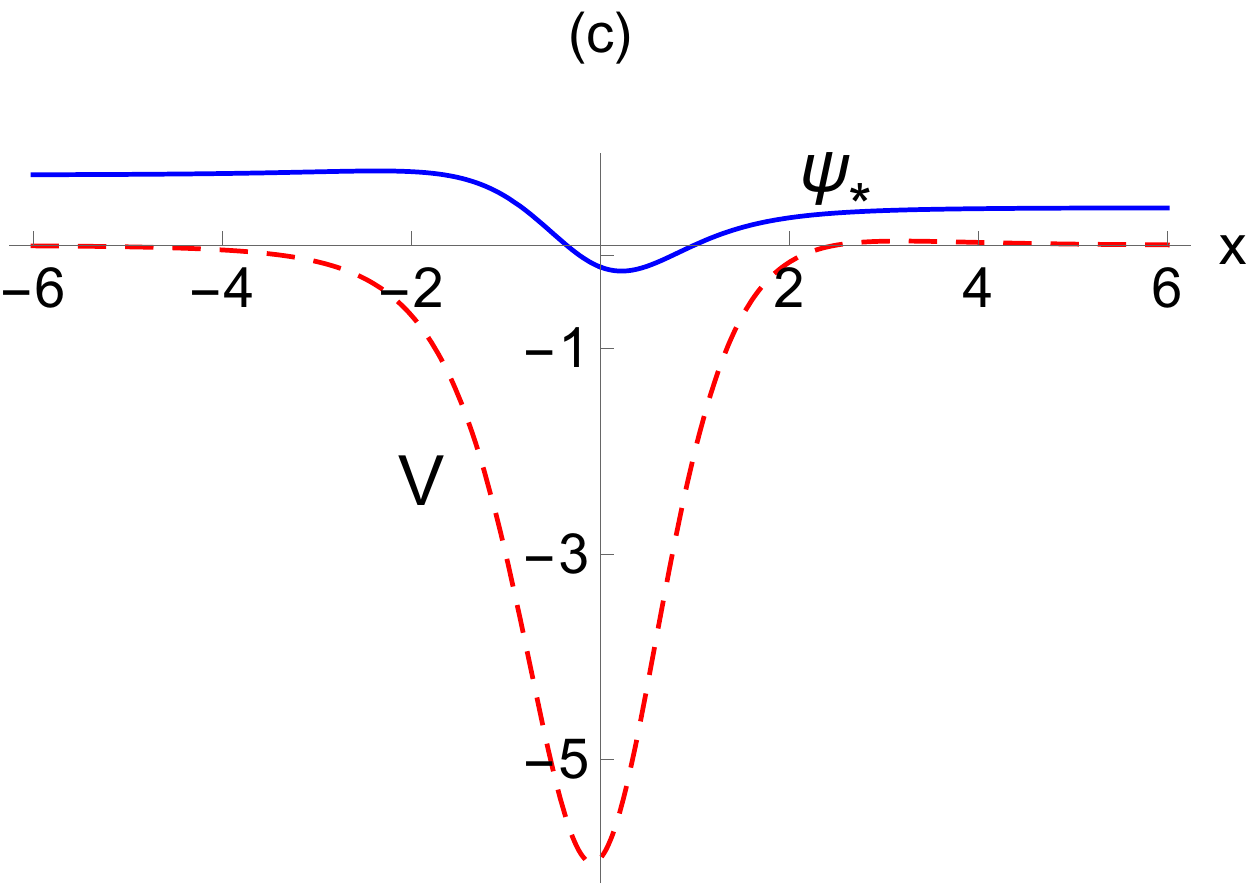}
	%\hskip .5 cm
	\caption{Half bound states in three cases of Scarf II potential (5), when $s=0.2$;  (a): $q=0$, (b): $q=1$, (c): $q=2$, with zero, one and two node respectively. In each case $R(0)=\tanh^2(\pi s)=0.31013$. Notice that all three are well-barrier systems with barrier height decreasing from (a) to (c)
		for $x>0$. Interestingly, if $q$ is not an integer but close to an integer, $R(0)$ becomes 1, dramatically.} 
\end{figure}
\section{Two numerically solvable well-barrier systems}
\subsection{}
\begin{figure}[H]
	\centering
	\includegraphics[width=7 cm,height=5 cm]{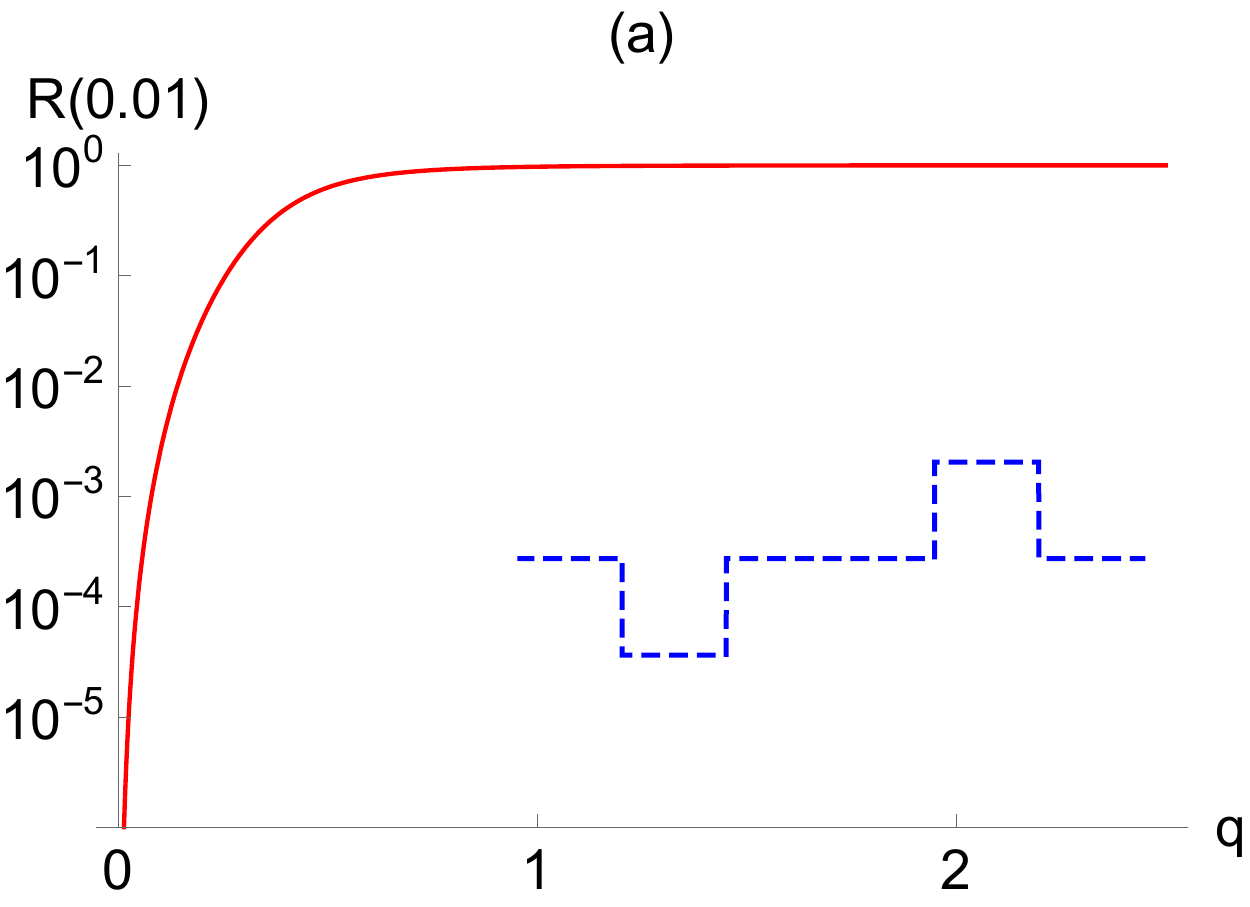}
	\hskip .5 cm
	%\includegraphics[width=4.5 cm,height=5 cm]{fig-2b.pdf}
	%\hskip .5 cm
	\includegraphics[width=7 cm,height=5 cm]{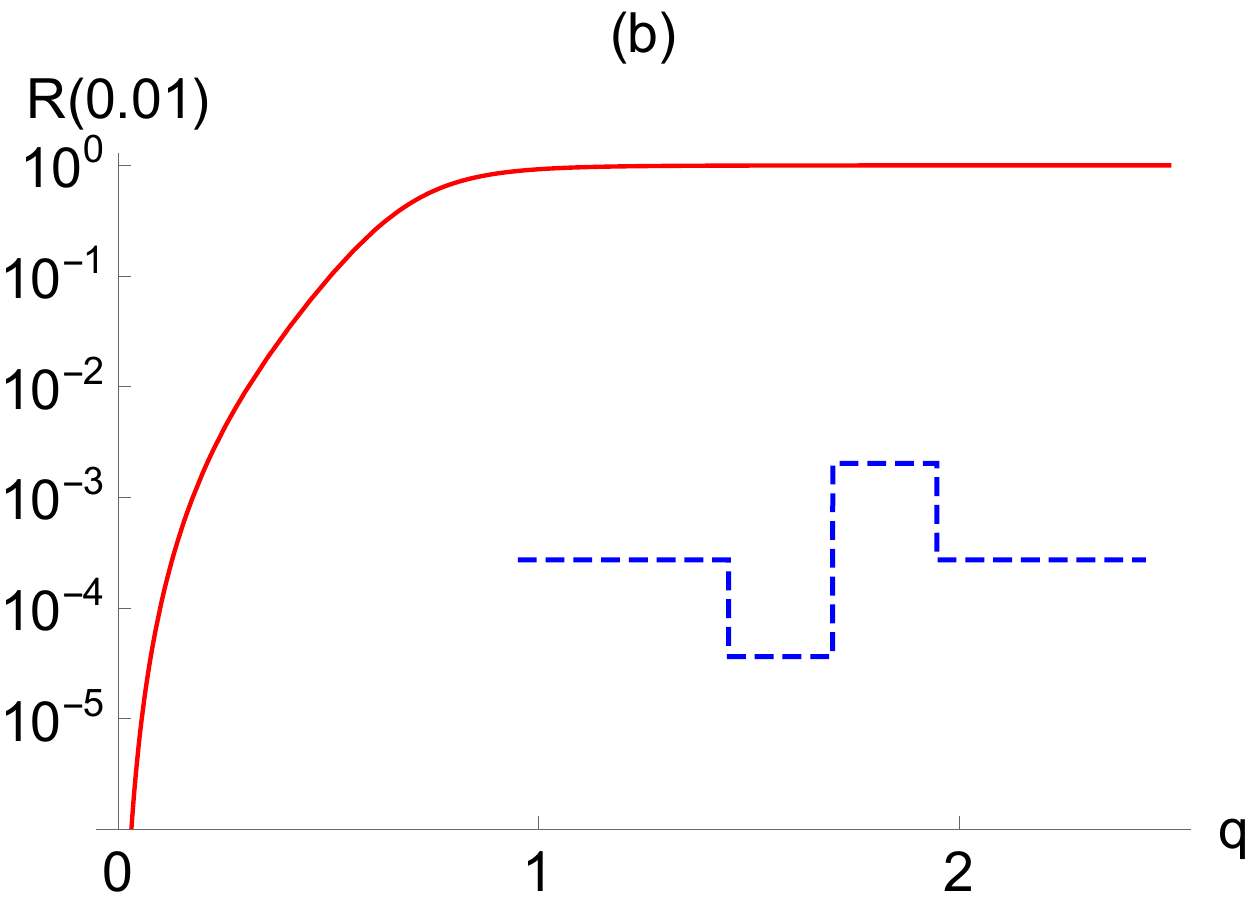}
	%\hskip .5 cm
	\caption{$R(0.01)$ (Solid) for anti-symmetric square well-barrier system (Dashed) when $u_1=u_2=u_0$ and the distance $a$ between well and barrier is varying as 2, 0 for (a), (b) repectively. Due to anti-symmetry there is a band of low $q$ values ($q<0.5)$ and nowhere else where the $R(0.01)<<1$ and when this distance decreases the band of $q$ values increases. Here, $q=w \sqrt{u_0}$ and $w$ is the fixed width of the both well and barrier.}  
\end{figure}

\subsection{}
\begin{figure}[H]
	\centering
	\includegraphics[width=7 cm,height=5 cm]{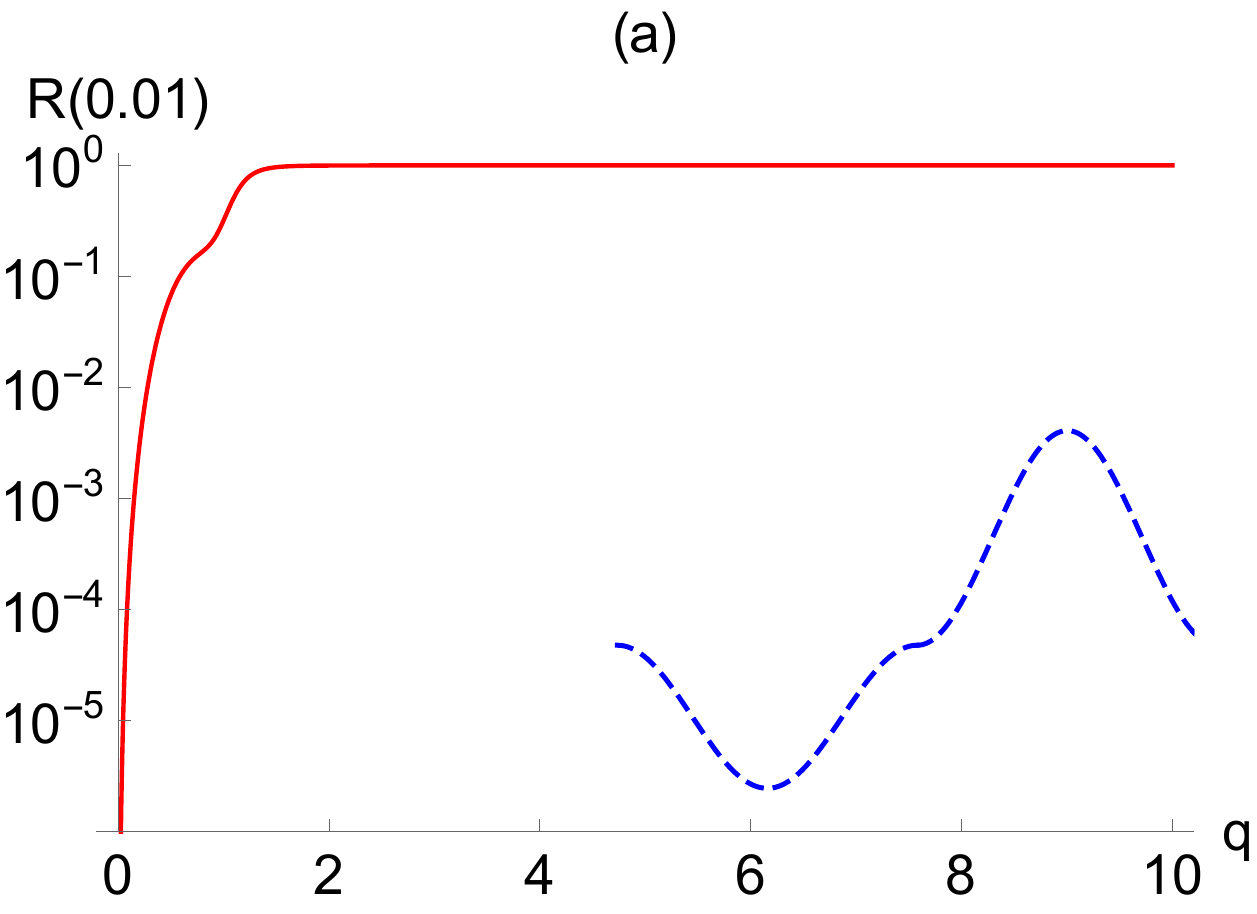}
	\hskip .5 cm
	%	\includegraphics[width=4.5 cm,height=5 cm]{fig-3b.pdf}
	%	\hskip .5 cm
	\includegraphics[width=7 cm,height=5 cm]{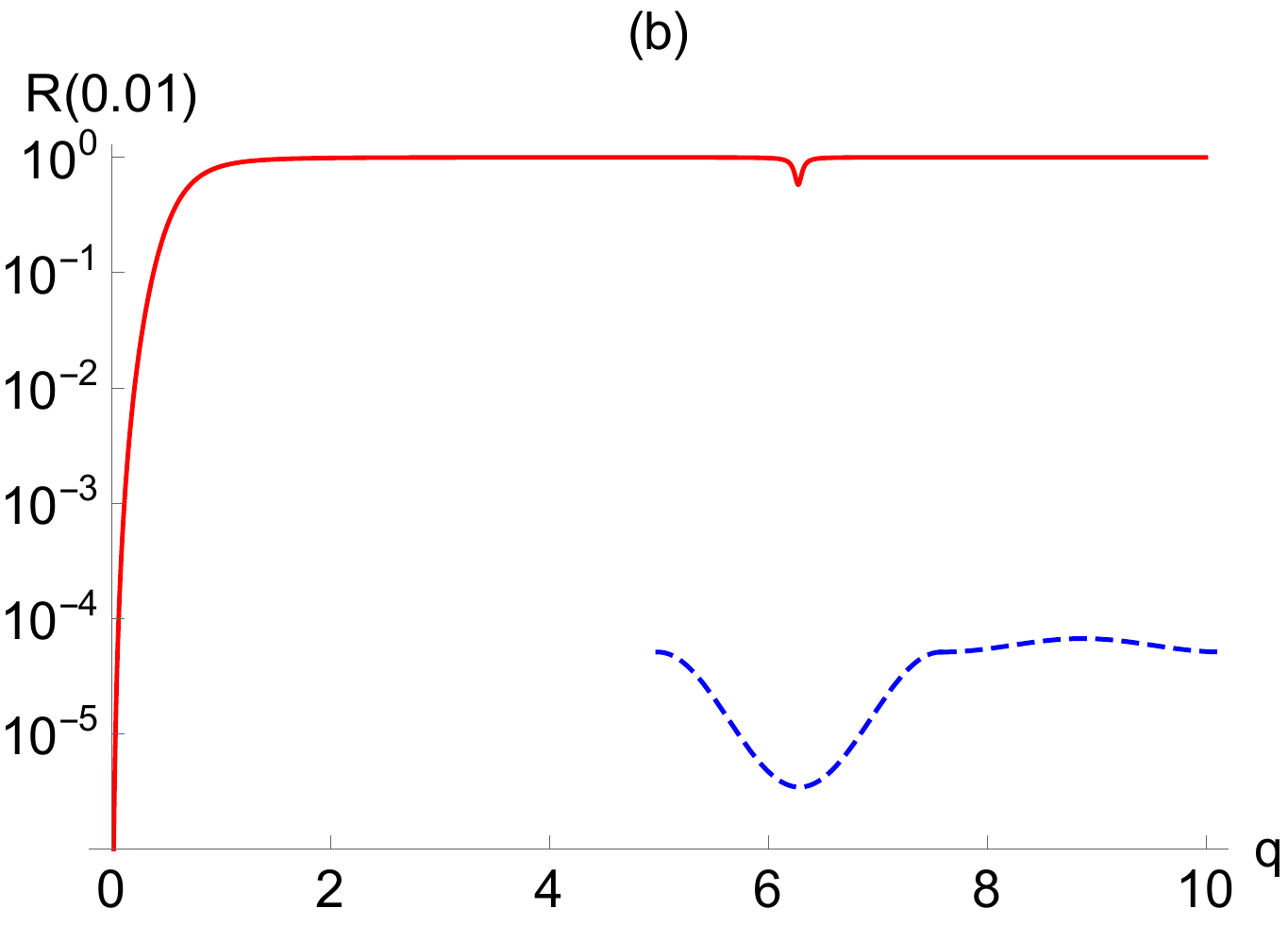}
	%\hskip .5 cm
	\caption{$R(0.01)$ (Solid) for asymmetric sin-squared  well-barrier system (Dashed), when $u_1=u_0=1$ and the barrier height($u_2=\eta u_0$) is reducing as, (a): $\eta=1.5$, (b): $\eta=0.1$. When barrier height decreases the band-width of small $q$ values($q <0.5$ in (a) and $q<1$ in (b)) reduces and for very small barrier height there will be low reflection around a large $q$ value (see (b)). Here $a=1$.} 
\end{figure}

We would like to conclude that we have presented two exactly and two numerically solved well-barrier scattering potentials which normally give
$R(0)=1$, however, when their parameters attain critical values, we observe: $R(0)<1$ or even $R(0)<<1$. In these cases, the potential possesses $E=0$ as $n$-node half bound states ($n=0,1,2,...$), consequently there are $n$ number of bound states. Also in these cases, we get low reflection at low energies which is paradoxical. Quantum tunnelling is classically non-intuitive, the low reflection at low energies is even more so.

\end{document}